\documentclass[11pt,a4paper]{article}
\pdfoutput=1

\usepackage{jcappub}

\usepackage{epsf,epsfig}
\usepackage{graphics}
\usepackage{color}
\usepackage{cancel} 
\usepackage{ulem}

\newcommand{\beq}{\begin{equation}}
\newcommand{\eeq}{\end{equation}}
\newcommand{\bea}{\begin{eqnarray}}
\newcommand{\eea}{\end{eqnarray}}

\newcommand{\bmatr}{\begin{pmatrix}}
\newcommand{\ematr}{\end{pmatrix}}

\title{Inflaton perturbations through an Ultra-slow-roll transition and Hamilton-Jacobi attractors}

\author{Tomislav Prokopec$^a$ and Gerasimos Rigopoulos$^b$}
\affiliation{$^a$  Institute for Theoretical Physics, Spinoza Institute \& {\rm EMME}$\Phi$, Utrecht University, Princetonplein 5, 3584 CC Utrecht, The Netherlands \\
$^b$ School of Mathematics, Statistics and Physics, Herschel Building, Newcastle University, Newcastle upon Tyne, NE1 7RU, UK
}

\abstract{
We examine the behaviour of the gauge invariant scalar field perturbations in an analytic inflationary model that transitions from slow-roll to an ultra-slow-roll (USR) phase. We find that the numerical solution of the Mukhanov-Sasaki equation is well described by Hamilton-Jacobi (HJ) theory, as long as the appropriate branches of the Hamilton-Jacobi solutions are invoked: Modes that exit the horizon during the slow-roll phase evolve into the USR as described by the first HJ branch, up to a subdominant $\mathcal{O}(k^2/H^2)$ correction to the Hamilton-Jacobi prediction for their final amplitude that we compute, indicating the influence of neglected gradient terms. Modes that exit during the USR phase  are described by a separate HJ branch once they become sufficiently superhorizon, obtained by the {shift $\left(\epsilon_1,\epsilon_2\right) \simeq \left(0,-6+\Delta \right) \rightarrow \left(\tilde{\epsilon}_1,\tilde{\epsilon}_2\right)\simeq (0,-\Delta)$ and corresponding to a slow-roll solution (very close to de Sitter) supported by the same potential. This transition is similar to the conveyor belt concept put forward in our previous work~\cite{Prokopec:2019srf} and suggests that the limit $\epsilon_2\rightarrow -6$ is unphysical as an asymptotic value for the background/long wavelength solution.} We further discuss implications for the validity of the stochastic equations arising from the Hamilton-Jacobi formulation. {Our work suggests that if Hamilton-Jacobi attractors are appropriately used, they can successfully describe the dynamics of long wavelength inflationary inhomogeneities for potentials with USR regions.}
}

\begin{document}
\maketitle

\section{Introduction}
\label{sec:intro}

In this paper we examine the application of the Hamilton-Jacobi (HJ) approach for describing the dynamics of inhomogeneous fields in an inflationary scenario where a slow-roll (SR) phase transitions to an ultra-slow-roll (USR) phase. We find in our model that gauge invariant perturbations are eventually described by a HJ branch that is {\it separate to the one they start on}, where field perturbations practically fossilize. The initial SR phase is important as it generates cosmological perturbations that induce cosmic microwave background (CMB) fluctuations and seed Universe's structure formation. A USR inflationary phase~\cite{Tsamis:2003px,Kinney:2005vj} 
is of essential importance for post-inflationary production of primordial black holes (PBHs)~\cite{Germani:2017bcs}, which are a viable dark matter candidate~\cite{Carr:2021bzv,Escriva:2022duf,LISACosmologyWorkingGroup:2023njw}, and which can be produced in a variety of inflationary 
models~\cite{Kawasaki:2016pql,Domcke:2017fix,Ezquiaga:2017fvi, Rigopoulos:2021nhv}. Furthermore, understanding stochastic dynamics in the context of USR inflation is important, as it can significantly affect production of 
PBHs~\cite{Pattison:2017mbe,Ezquiaga:2019ftu,Pattison:2021oen,Animali:2022otk}.

To understand USR and the associated dynamics of perturbations, we studied in \cite{Prokopec:2019srf} an analytically tractable model of the inflationary dynamics and long wavelength fluctuations of a scalar field sliding on a constant $V=V_0$ potential, using the HJ formalism \cite{Salopek:1990jq}. We argued there that  the system evolves from one HJ attractor, on which the long wavelength field slows down {(but always with $\dot{\phi}\neq 0$)} and field inhomogeneities decay (all spatial points asymptotically approaching a point $\phi_0$ the value of which depends on the initial velocity), to a second HJ attractor without classical velocity { ($\dot{\phi} = 0$)} where the field freely diffuses on the $V=V_0$ surface, which is simply the dynamics of a free massless scalar in de Sitter. An important point of the argument in \cite{Prokopec:2019srf} was that both decaying and constant $\phi$ behaviours were solutions of the HJ equation and both satisfied the momentum constraint in the long wavelength limit implicit in the HJ formalism. We used this idealized system to examine the transition to de Sitter, also aiming to understand the "quantum diffusion" regime of USR potentials that exhibit a quantum well \cite{Pattison:2017mbe}, including the effect of the field's velocity. Since these two HJ branches did not evolve into each other by classical HJ dynamics, the transition between them was triggered by field fluctuations exiting the Hubble radius (modeled as stochastic noise) and was termed the ``conveyor belt'' in \cite{Prokopec:2019srf}.      

In this work we test this HJ conveyor belt notion by using an analytic toy model representing a more general inflationary evolution that leads from a SR into an USR phase. Our findings can be summarized as follows:

a) The amplitude of linear perturbation modes that cross the Hubble radius during the SR phase decay during USR, following the general prediction of the { initial} HJ attractor. Their behaviour eventually deviates form the HJ expectation of complete decay as these modes freeze out at a  residual subdominant amplitude which we find to be 
	\begin{equation}
		\frac{\delta\Phi(\alpha\rightarrow \infty)}{\delta\Phi_{\rm dS}}
		\simeq \mathcal{B} \left(\frac{k}{H}\right)^{\!2}
		\,.
		\label{USRL field decay}
	\end{equation}
Here, $\delta\Phi_{\rm dS}$ is the expected amplitude in an asymptotic de Sitter stage and  $\mathcal{B}$ is a constant depending on the model parameters. We obtain this $k^2/H^2$ dependence via analytical argument and extract $\mathcal{B}$ from the numerical solution of~(\ref{Field eq 1}). This residual field amplitude in USR is one of the principal results of this paper and shows that the HJ dynamics correctly carries pre-existing long wavelength modes into the USR phase, until neglected $k^2$ terms become important compared to the decaying perturbation amplitude {-- see \cite{Leach:2001zf} for an early work on the role of gradient corrections, and more recently \citep{Jackson:2023obv} and \citep{Raveendran:2025pnz}\footnote{Note that although the authors of~\citep{Raveendran:2025pnz} point out that gradient terms temporarily affect the evolution of perturbation modes and their velocity during the USR transition, no quantitative analysis is attempted there.}. }

b) Modes that exit during the USR phase are again described by an HJ attractor, after becoming sufficiently super-horizon, that is however {\it different} from the initial one: if during USR $(\epsilon_1, \epsilon_{2}) \rightarrow  (0,-6 + \Delta)$, where $\Delta\ll 1 $ depends on our model's slow-roll parameters, then the new HJ attractor is defined by $(\tilde{\epsilon}_1,\tilde{\epsilon}_2)=(0,-\Delta)$ and corresponds to a new HJ { (Slow-Roll)} solution around the asymptotic USR field value $\phi_0$ which is a critical point of our potential. {The behaviour of perturbations in this new branch is the same as in the initial branch in what is referred to in \cite{Karam:2022nym} as Wands duality \cite{Wands:1998yp}.\footnote{
The equivalence of the behaviour of perturbations for different cosmological backgrounds was noted in \cite{Wands:1998yp}. We note here however that the duality originally posed in \cite{Wands:1998yp} refers to a mapping of different complete cosmological expansion histories and not finite portions of them. In our case the mapping refers only to times when $\epsilon_1\simeq 0$ and $\epsilon_2\simeq-6$.}  
}    

c) The above implies that the long wavelegth dynamics of the inhomogeneous modes can in fact be described  via a transition between 2 otherwise disjoint HJ attractor branches, analogous to the conveyor belt idea discussed in  \cite{Prokopec:2019srf}. There, the decay of the field's velocity during its motion towards an asymptotic point with $\dot{\phi}\neq 0$ was followed by a transition to the de Sitter solution with constant $\phi$, thereby joining the two distinct HJ branches of the $V=V_0$ potential, {both branches being solutions of \eqref{HJ} and \eqref{momentum-def} for the $V=V_0$ potential}. Our results here show that this idea applies to the model studied here as well and hints to its relevance to more general scenarios. An extra ingredient, uncovered here, are the subdominant $k^2/H^2$ terms which dictate a residual perturbation amplitude for those modes that exit the horizon prior to the USR phase.       

d) Our analysis supports the use of the HJ stochastic formalism for studying USR-type dynamics. This is achieved by the conveyor belt, i.e. appropriately joining the stochastic evolution two HJ attractor branches that align with the two linearly independent solutions of the perturbation modes. Our findings contradict the criticism of \cite{Artigas:2025nbm} which claims that using HJ in USR, as done in \cite{Prokopec:2019srf} and \cite{Rigopoulos:2021nhv}, is incorrect. By using different HJ branches the conveyor belt does not restrict the stochastic HJ equation to a subspace of the available phase space for the long wavelength field inhomogeneities, and this turns out to be important for applying HJ to USR.

We work in natural units, in which $\hbar = 1 = c$ and 
$M_{\rm P}^2 = 1/(8\pi G) = m_{\rm P}^2/(8\pi)=1$.

\section{A USR toy model}

\subsection{The background model}

\begin{figure}[t]
	\centering
	\includegraphics[width=0.45\textwidth]{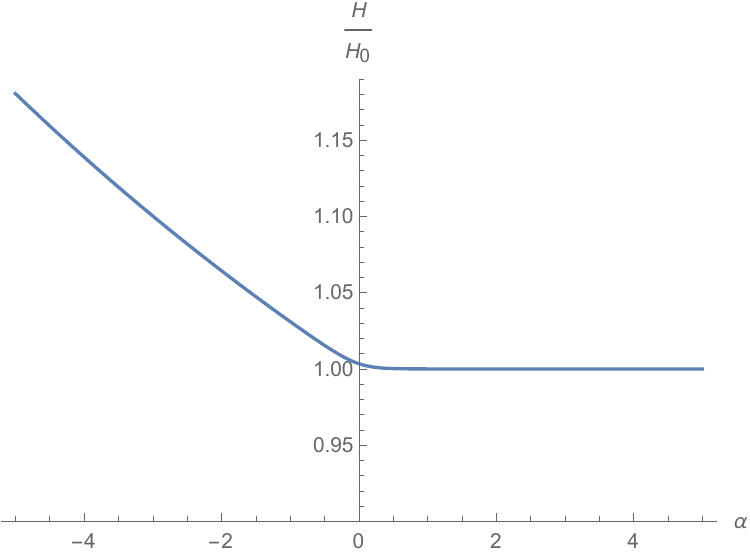}
    \hskip 1cm \includegraphics[width=0.45\textwidth]{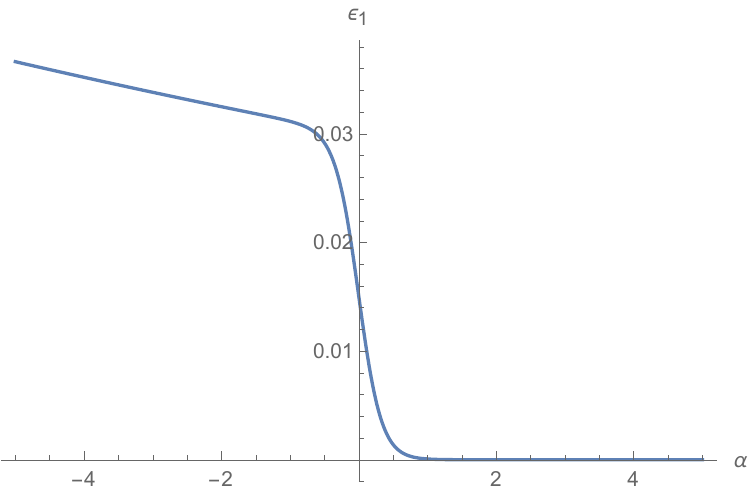}
    \includegraphics[width=0.45\textwidth]{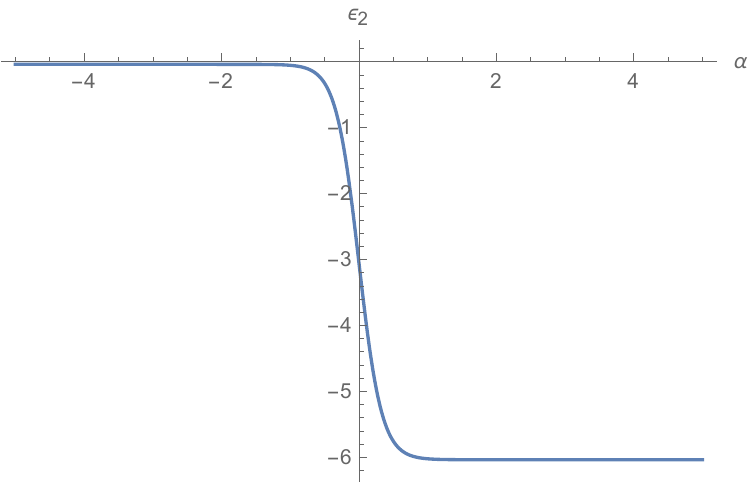}
    \hskip 1cm \includegraphics[width=0.45\textwidth]{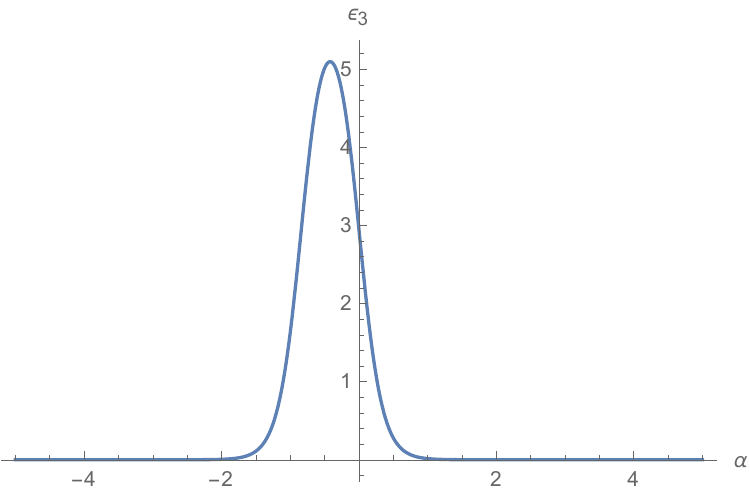}
	\caption{USR evolution {in our model  \eqref{appx: SR to USR} with $\epsilon_0 = 0.03$ and $\eta_0 = -0.02$, used in the numerical examples of this paper}: Top panel (left to right): $H(\alpha)/H_0$, $\epsilon_1(\alpha)$. Bottom panel (left to right): $\epsilon_2(\alpha)$, $\epsilon_3(\alpha)$.}
	\label{fig:USR-evol}
\end{figure}

We consider the following model to describe a transition from an initial SR phase to a period of USR. Instead of { starting with} defining a potential, we work directly with the HJ slow-roll parameters, taking the first slow-roll parameter to be 
\begin{equation}
	\epsilon \equiv \epsilon_1 
	= \frac{\epsilon_0{\rm e}^{2\eta_0 \alpha}}{1+{\rm e}^{6\alpha}}
	\label{appx: SR to USR}
\,,\quad
\end{equation}
{ where $\alpha = \ln(a)$ is a time variable parametrizing inflation e-foldings} and we take $\epsilon_0\ll 1$ and $\|\eta_0\|\ll 1$ with $\eta_0 < 0$ or $\eta_0 > 0$. The higher order slow-roll parameters (see appendix \ref{Appendix A} for definitions) follow:
\begin{equation}
	\epsilon_2 = 2\eta_0 - \frac{6}{1+{\rm e}^{-6\alpha}}
	\,,\qquad \epsilon_3 
	= \frac{18}{\left[\big(3- \eta_0\big){\rm e}^{6\alpha}-\eta_0 \right]}
	\frac{1}{1+{\rm e}^{-6\alpha}}
	\,.\quad
	\label{appx: eps2 and eps3}
\end{equation}
The Hubble parameter can be obtained by integrating $\epsilon_1$ 
from $+\infty$ to $\alpha>0$,~\footnote{It is convenient to pick one limit of integration
in Eq.~(\ref{appx: Hubble parameter}) to be $\alpha=+\infty$, because $H$ asymptotes to a constant irrespective of the signs of $\eta_0$ and $\epsilon_0$.}
\begin{eqnarray}
	\ln\left(\frac{H(\alpha)}{H_0}\right) &=& -\int_{\infty}^\alpha \epsilon_1(\alpha'){\rm d}\alpha'
	= - \frac{\epsilon_0}{6}
	{\rm e}^{2\eta_0\alpha}
	\sum_{n=1}^\infty (-1)^n\frac{{\rm e}^{-6n\alpha}}{n-\frac{\eta_0}{3}}
	\nonumber\\
	&=& \frac{\epsilon_0}{2\eta_0}{\rm e}^{2\eta_0\alpha}
	\times\Big[{}_2F_1\Big(1,-\frac{\eta_0}{3};1-\frac{\eta_0}{3};
	-{\rm e}^{-6\alpha}\Big)-1\Big]
	\,,
	\label{appx: Hubble parameter}
\end{eqnarray}
where $H_0=H(\infty)$.
%
%
Eq.~(\ref{appx: Hubble parameter}) is suitable for calculating 
the expansion rate in USR when $\alpha>0$. If one is interested in the expansion rate during the preceding slow-roll phase ($\alpha<0$),
a more suitable expression can be obtained by 
making use of Gauss' relation~(9.132.2) in Ref.~\cite{Gradshteyn:1943cpj}
(which relates hypergeometric functions of $z$ to the ones of $1/z$),
in which case 
Eq.~(\ref{appx: Hubble parameter}) can be also written as,
\begin{eqnarray}
	\ln\left(\frac{H(\alpha)}{H_0}\right)
	&=& \frac{\epsilon_0}{2\eta_0}
	\times\Bigg[\Gamma\left(1\!+\!\frac{\eta_0}{3}\right)
	\Gamma\left(1\!-\frac{\eta_0}{3}\right)
    -{\rm e}^{2\eta_0\alpha}
	\nonumber\\
	&& +\, \frac{\eta_0}
	{3+\eta_0}
	{\rm e}^{2(3+\eta_0)\alpha}
	{}_2F_1\Big(1,1\!+\!\frac{\eta_0}{3};2\!+\!\frac{\eta_0}{3};
	\!-{\rm e}^{6\alpha}\Big)
	\Bigg]
	\,,\quad
	\label{appx: Hubble parameter 2}
\end{eqnarray}
which in the limit when $\alpha\rightarrow -\infty$ reduces to,
\begin{eqnarray}
	\ln\left(\frac{H(\alpha)}{H_0}\right)
	&=& \frac{\epsilon_0}{2\eta_0}
	\times\bigg[\Gamma\left(1\!+\!\frac{\eta_0}{3}\right)
	\Gamma\left(1\!-\frac{\eta_0}{3}\right)\!-\!{\rm e}^{2\eta_0\alpha} 
	\bigg]
	\,,\quad
	\label{appx: Hubble parameter 3}
\end{eqnarray}
such that at early times ($\alpha\rightarrow -\infty$)
the expansion rate asymptotes to a constant 
(when $\eta_0>0$), or grows exponentially with $\alpha$ (when $\eta_0<0$). These functions are plotted in Fig.~\ref{fig:USR-evol}.

In the limit $\eta_0 \rightarrow 0$ one obtains a simpler model in which the system transits from 
a {\it constant roll} $\epsilon\simeq \epsilon_0 ={\rm const.}$ into an USR phase:
\begin{equation}
	\epsilon_1(\alpha) = \frac{\epsilon_0}{1+{\rm e}^{6\alpha}}
	\,,\qquad 
	\epsilon_2(\alpha) = -\frac{6}{1+{\rm e}^{-6\alpha}}
	\,,\qquad
	\epsilon_3(\alpha) = \frac{6}{1+{\rm e}^{6\alpha}}
	\,,
	\qquad 
\label{simple model USR}
\end{equation}
in which $\epsilon_1$ rapidly decays and the Hubble rate is 
\begin{equation}
H(\alpha) = H_0\left(1+{\rm e}^{-6\alpha}\right)^\frac{\epsilon_0}{6}
\,.\quad
\end{equation}
This limiting case is more like our earlier study of a slide on the  $V=V_0$ potential where the system enters a pure de Sitter phase. In this paper we will focus on the $\eta_0 < 0$ case for our numerical example, taking $\epsilon_0 = 0.03$ and $\eta_0 = -0.02$. 

Given $H(\alpha)$ one can explicitly compute a potential function $V(\phi)$ that would give rise to it. For the simple model~(\ref{simple model USR}) we have,
\begin{equation}
\partial_\alpha\phi = \sqrt{2\epsilon} \;\Longrightarrow\; 
  \phi(\alpha) = \phi_0 - \frac{\sqrt{2\epsilon_0}}{3}{\rm Arcsinh}\left({\rm e}^{-3\alpha}\right)
\,,\qquad
\label{simple model: phi of alpha}
\end{equation}
from where,
\begin{equation}
V(\phi) = (3-\epsilon)H^2 = H_0^2
   \left[3-\epsilon_0\tanh^2\left(\frac{3(\phi-\phi_0)}{\sqrt{2\epsilon_0}}\right)\right]
    \left[\cosh\left(\frac{3(\phi-\phi_0)}{\sqrt{2\epsilon_0}}\right)\right]^\frac{2\epsilon_0}{3}
\,.\qquad
\label{simple model: V of phi}
\end{equation}
Inflation in this model proceeds from a large field value $\phi\ll \phi_0$. During the early stages of inflation the inflaton rolls in a slow-roll attractor, reaching eventually an USR regime, in which the field approaches $\phi_0$ as $\alpha\rightarrow\infty$. To get a graceful exit in this model, one would have to glue another potential at $\phi\approx \phi_0$, which would provide a second period of inflation and/or an exit from inflation. This model closely resembles the USR model discussed in \cite{Prokopec:2019srf},
where the conveyor belt concept was introduced.

\begin{figure}[t]
	\centering
    \vskip -0.3cm
	\includegraphics[width=0.45\textwidth]{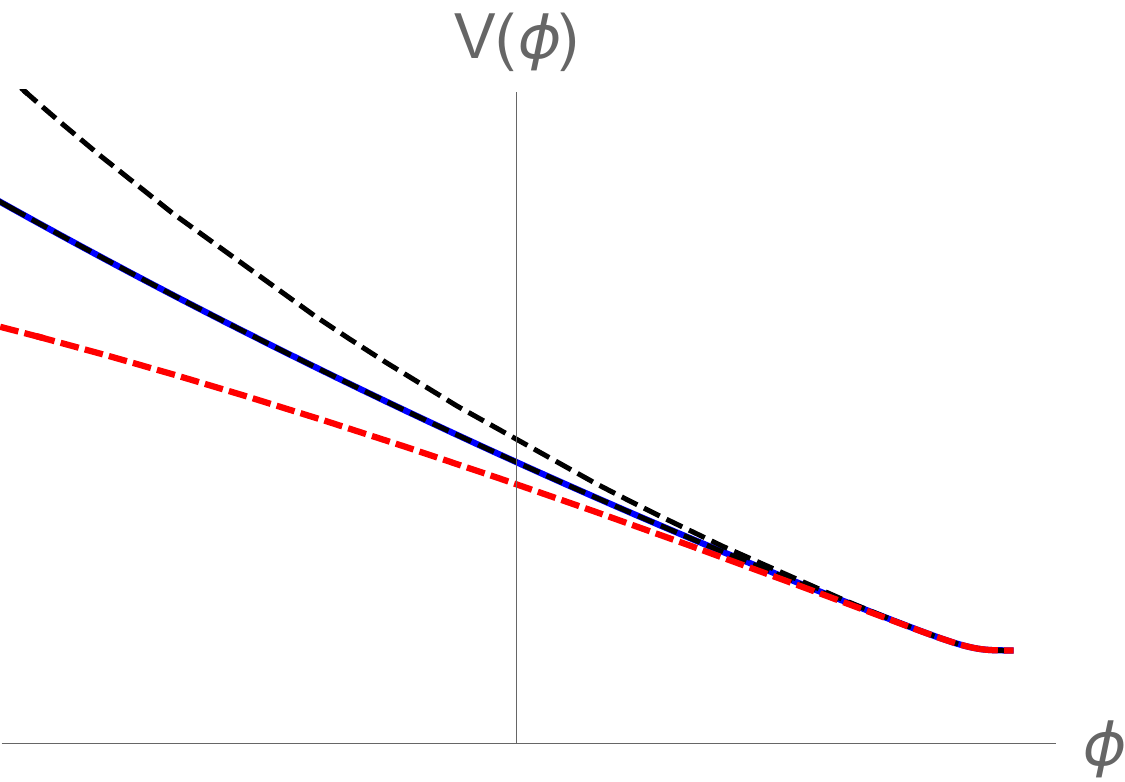}\hskip 1cm
    \includegraphics[width=0.45\textwidth]{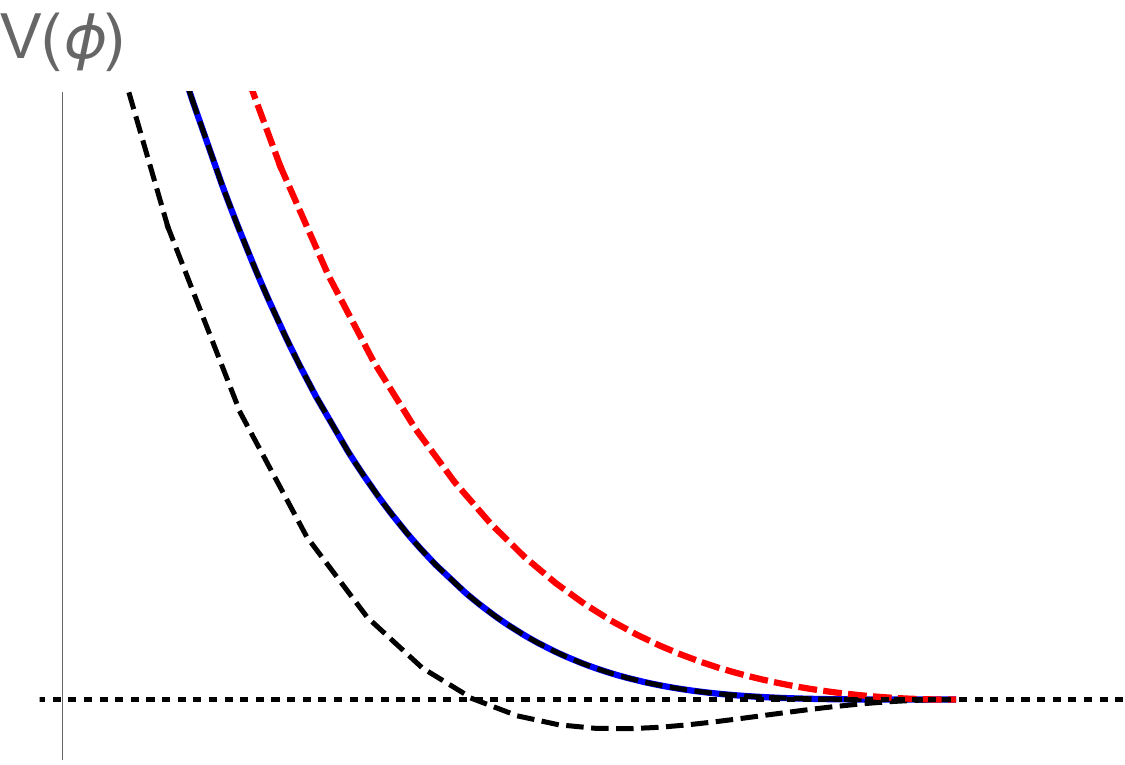}
	\caption{ {\it Left panel:} The simple potential in Eq.~(\ref{simple model: V of phi}) (solid blue)
    and the more realistic one with $\eta_0>0$ (dashed red) and $\eta_0<0$ (dashed black), given in parametric form in Eqs.~(\ref{realistic model: phi of alpha})--(\ref{realistic model: V of alpha}). {\it Right panel:} Near the minimum both potentials exhibit a rather flat section, in which the USR period occurs. A zoomed-in image of the potential reveals a gentle slope around $\phi_0$ caused by $\eta_0\neq 0$. The potential can be continued past the asymptotic field value $\phi_0$, for example by gluing a constant $V=V_0$ part, or a symmetric image - see \eqref{potential around minimum}.}
	\label{fig:Potential}
\end{figure}

There is no simple way of constructing the inflationary potential for the more complicated, but more realistic model in Eqs.~(\ref{appx: SR to USR})--(\ref{appx: eps2 and eps3}).
Nevertheless, one can construct a parametric form of the potential.
Integrating $\partial_\alpha\phi = \sqrt{2\epsilon}$
gives ({\it cf.} Eq.~(\ref{simple model: phi of alpha})),~\footnote{
By recalling that, 
\begin{equation}
{}_2F_1\left(\frac{1}{2},\frac{1}{2};\frac{3}{2};
	-{\rm e}^{-6\alpha}\right) = {\rm e}^{3\alpha}{\rm Arcsinh}\Big({\rm e}^{-3\alpha}\Big)
\,,\qquad
\nonumber
\end{equation}
one can easily see that,
in the limit when $\eta_0\rightarrow 0$, Eq.~(\ref{realistic model: phi of alpha}) reduces to 
Eq.~(\ref{simple model: phi of alpha}).
}
\begin{equation} 
  \phi(\alpha) = \phi_0 - \frac{\sqrt{2\epsilon_0}}{3-\eta_0}
  \!\times\!{\rm e}^{-(3-\eta_0)\alpha}{}_2F_1\left(\frac{1}{2},\frac{1}{2}-\frac{\eta_0}{6};\frac{3}{2}-\frac{\eta_0}{6};
	-{\rm e}^{-6\alpha}\right)
\,,\qquad
\label{realistic model: phi of alpha}
\end{equation}
and the potential $V(\phi)$ can be also written as a function of $\alpha$
 ({\it cf.} Eq.~(\ref{simple model: V of phi})),
\begin{equation}
V(\phi) = \big(3-\epsilon(\alpha)\big)H^2(\alpha) 
\,,\qquad
\label{realistic model: V of alpha}
\end{equation}
where $\epsilon(\alpha)$ and $H^2(\alpha)$ are given in Eqs.~(\ref{appx: SR to USR})
and~(\ref{appx: Hubble parameter})--(\ref{appx: Hubble parameter 2}), respectively.
The potentials~(\ref{simple model: V of phi}) 
and~(\ref{realistic model: phi of alpha})--(\ref{realistic model: V of alpha}) are plotted in Figure~\ref{fig:Potential}.
The model~\eqref{appx: SR to USR} corresponds to a potential with a slow-roll supporting part leading to a shallow minimum
visible in Figure~\ref{fig:Potential}. The relevant 
part of the potential around $\phi=\phi_0$ is,
\begin{equation} 
    V(\phi) \simeq 3H_0^2\left[1
    +\frac{(3\!-\!\eta_0)\eta_0}{6}\big(\phi\!-\!\phi_0)^2
    +\left(\frac{9(3\!-\!\epsilon_0)}{8\epsilon_0}
    +\mathcal{O}(\eta_0)\right)
        \big(\phi\!-\!\phi_0)^4\right]
\,,
\label{potential around minimum}
\end{equation}
from where we see that for $\eta_0<0$ 
the potential exhibits a double minimum at 
$\phi=\phi_0 \pm\sqrt{-\frac{2\epsilon_0\eta_0(3\!-\!\eta_0)}{27(3\!-\!\epsilon_0)}}$ 
(when $\eta_0>0$ the minimum remains at $\phi=\phi_0$). This means that, during the USR, the field climbs a gentle hill while slowing down as the spacetime approaches de Sitter, asymptotically approaching the shallow peak at $\phi=\phi_0$ as $\alpha\rightarrow \infty$. 

{ It is important to note here that the potential \eqref{potential around minimum} that supports \eqref{appx: SR to USR} can also support a slow roll evolution, {\it separate} to the history defined by \eqref{appx: SR to USR} and \eqref{appx: eps2 and eps3}. This can be seen by setting $H \simeq \sqrt{V/3}$ and using \eqref{eq:eps2} to compute $\epsilon_2$, giving 
\begin{equation}\label{Veps2}
\frac{\epsilon_2}{2}\simeq 2\left(\left(\frac{\partial_\phi V}{V}\right)^2- \frac{\partial_{\phi\phi} V}{V}\right)\,.
\end{equation}
For the potential \eqref{potential around minimum} we have $\epsilon_2 \rightarrow -2\eta_0$ and $\epsilon_1\simeq 0$ for $\phi\rightarrow\phi_0$. This separate cosmological branch will be very important for our argument below.  } 

\subsection{Perturbations}

Linearized perturbations obey second order differential equations, with scalar perturbations being concisely described by the Mukhanov-Sasaki equation,
\begin{equation}
	\Big(\partial_\tau^2 + k^2 - \frac{z''}{z}\Big)Q(\tau,\vec k) = 0
	\,,
	\label{Mukhanov eq}
\end{equation}
where $Q = a\delta \phi-z\psi$ 
is the gauge invariant Mukhanov-Sasaki field (in terms of which the gauge invariant curvature perturbation is given by 
${\cal R}=-Q/z$), $\psi$ is the zero-order-in-gradients scalar perturbation of the spatial metric (the other one being usually denoted by $\partial_i\partial_jE$), { $\tau$ is conformal time: $a{\rm d}\tau=N{\rm d}t$, $z = \sqrt{2\epsilon_1}a$ and $z'\equiv \partial_\tau z$.} To convert this equation to the formalism used here, first notice that $\partial_\tau = \frac{a}{N}\partial_t = H{\rm e}^\alpha\partial_\alpha$.
This then implies for $z$, 
\begin{equation}
	z'(\tau) = aH\left(1+\frac{\epsilon_1}{2}\right)z
	\label{appx: z prime}
\end{equation}
and 
\begin{eqnarray}
	\frac{z''}{z} &=& (aH)^2\left[2-\epsilon_1+\frac32\epsilon_2
	-\frac12\epsilon_1\epsilon_2
	+\frac14\epsilon_2^2
	+\frac12\epsilon_2\epsilon_3\right]
%
\,.
	\label{appx: z'' over z}
\end{eqnarray}
Inserting these results and using
\begin{equation}
	\partial_\tau^2 
	= (Ha)^2\big[(1-\epsilon)\partial_\alpha +\partial_\alpha^2\big]
	\,,\qquad
	\label{appx: conversion d eta}
\end{equation}
into 
Eq.~(\ref{Mukhanov eq}) one obtains, 
\begin{eqnarray}
	\left\{\partial_\alpha^2 + (1-\epsilon_1)\partial_\alpha+ \frac{k^2}{(aH)^2}
	- \left[2-\epsilon_1+\frac32\epsilon_2
	-\frac12\epsilon_1\epsilon_2
	+\frac14\epsilon_2^2
	+\frac12\epsilon_2\epsilon_3\right]
	\right\}Q(\alpha;\vec k) &=& 0
    \,,
	\qquad\;\;
	\label{Mukhanov eq 2}
\end{eqnarray}
or, for the (gauge invariant) field perturbation, ${\delta\Phi}=Q/a$
\begin{eqnarray}
	\left\{\partial_\alpha^2 + (3-\epsilon_1)\partial_\alpha+ \frac{k^2}{(aH)^2}
	+m^2(\alpha)\right\}\delta\Phi(\alpha; \vec{k}) &=& 0
    \,,
	\qquad
	\label{Field eq 1}
\end{eqnarray}
where the mass term for 
$\delta \Phi$ in Eq.~(\ref{Field eq 1}) is
\begin{equation}
	m^2(\alpha) \equiv  -\frac32\epsilon_2
	+\frac12\epsilon_1\epsilon_2
	-\frac14\epsilon_2^2
	-\frac12\epsilon_2\epsilon_3
	\,.
	\label{mass term of Q}
\end{equation} 
For what follows, it is important to note that as $\epsilon_1\rightarrow 0$ and $\epsilon_3\rightarrow 0$, the mass term \eqref{mass term of Q} gives $m^2\rightarrow \left(3-\eta_0\right)\eta_0$ for both $\epsilon_2 \rightarrow -6+2\eta_0$ which is the asymptotic endpoint of the USR evolution, and $\tilde{\epsilon}_2 \rightarrow -2\eta_0$ which corresponds to a new HJ branch, { a slow-roll solution for the potential \eqref{potential around minimum} around $\phi_0$ - see discussion around \eqref{Veps2}}. Hence, perturbation modes do not distinguish between backgrounds characterized by $\epsilon_2$ and $\tilde{\epsilon}_2$ when $\epsilon_1\rightarrow 0$ and $\epsilon_3\rightarrow 0$. As will be demonstrated by our numerical results below, it is the HJ branch with $\tilde{\epsilon}_2$ that is the correct background to use when $\Pi\rightarrow 0$. This is in accordance to the conveyor belt concept put forward in \cite{Prokopec:2019srf} and is generalized here beyond the $V=V_0$ potential.

The following relation can also be demonstrated (see e.g.~Eq.~(8.57) in Ref.~\cite{Mukhanov:2005sc})
%
\begin{eqnarray}
\left[\partial_\alpha -\frac{\epsilon_2}{2}\right]\delta\Phi
= - \sqrt{\frac{\epsilon_1}{2}}\frac{k^2}{(aH)^2}\Psi
\,,\quad
\label{relation to k2}
\end{eqnarray}
where $\Psi$ is the corresponding gauge invariant Bardeen potential associated with $\psi$: $\Psi=\psi+\frac{a'}{a}\left(B-E'\right)$ and $B$ is the scalar metric perturbation $g_{0i} = \partial_iB$. When the $k^2$ terms are ignored, we have that  
\beq\label{eq:long_modes}
\partial_\alpha\delta{\Phi}_k \simeq \frac{\epsilon_2}{2} \delta\Phi_k
\,,\qquad
\eeq   
{ or with $\epsilon_2\rightarrow\tilde{\epsilon}_2$ depending on the HJ branch used.} Hence, if the long wavelength approximation is valid we expect \eqref{eq:long_modes} to apply, {either with $\epsilon_2$ or $\tilde{\epsilon}_2$}. As we will see, using HJ for describing long wavelength dynamics also leads to \eqref{eq:long_modes}. 

We examine the validity of \eqref{eq:long_modes} below for our USR model, using numerical solutions of the Sasaki-Mukhanov equation { which exhibit the transition between the two branches. This transition is not part of the HJ equations \eqref{HJ} and \eqref{momentum-def}, indicating the influence of neglected gradient terms during USR. As demonstrated in \cite{Artigas:2025nbm}, such terms can indeed become relevant when the USR phase is reached. However, unlike the conclusion of that work we suggest that this does not invalidate the HJ framework, with these terms having only a transient effect, pushing instead the inhomogeneous system to a different solution of the HJ equations for the same potential. }

\section{Hamilton-Jacobi (Stochastic) dynamics}\label{sec:Stoch-HJ}

The long wavelength approximation, first advocated by Salopek and Bond \cite{Salopek:1990re,Salopek:1990jq} and used in \cite{Prokopec:2019srf}, leads to the Hamilton-Jacobi equation,
\beq\label{HJ}
\left(\frac{dH}{d\phi}\right)^2 = \frac{3}{2} H^2 -\frac{1}{2}V(\phi)
\,,\quad
\eeq
solutions of which are functions $H(\phi)$. We further note the definitions: 
\beq
H=\frac{1}{N}\frac{\partial \ln a}{\partial t}\,,\quad \Pi=\frac{1}{N}\frac{\partial \phi}{\partial t}\,,\quad 
\epsilon = \frac{1}{2}\frac{\Pi^2}{H^2}
\,,\quad \eta=-3-\frac{V'}{H\Pi}=
\frac{1}{N}\frac{\partial_t\Pi}{H\Pi}
\,.\quad
\eeq
Ignoring the l.h.s.~of \eqref{HJ} leads to the familiar slow-roll approximation. The momentum $\Pi$ is determined by the field $\phi$ via
\beq\label{momentum-def}
\Pi = -2\frac{dH}{d\phi}
\,,\quad
\eeq 
given a solution to (\ref{HJ}) and thus the only dynamical equation for the field in the HJ formalism is 
\beq\label{eq:evol-HJ}
\frac{1}{N}\frac{d\phi}{dt} = -2\frac{dH}{d\phi}
\,.
\eeq 
The HJ formulation neglects second order gradient terms and as a consequence also neglects the anisotropic part of the extrinsic curvature which is generally assumed to be exponentially decaying. 

We are interested here in studying the implications of assuming the validity of HJ dynamics for perturbations and stochastic evolution. These can be treated concurrently and we therefore add a Gaussian stochastic noise component to the dynamics
\beq\label{HJ-Stoch}
\frac{1}{N}\frac{d\phi}{dt} = -2\frac{dH}{d\phi} + \xi_\phi(t) 
\,,\quad
\eeq
satisfying 
\beq\label{noise1}
\left\langle \xi_\phi(t) \xi_\phi(t') \right\rangle =\mathcal{A}\delta(t-t')
\,.\quad
\eeq
The noise amplitude $\mathcal{A}$ need not be explicitly specified at this point but we can expect it to be of the order of $\mathcal{A}\simeq H_0^3/4\pi^2$ which is the result for a massless test scalar in de Sitter. We will however assume that $\mathcal{A}$ is {\it not} an explicit function of the stochastic variable $\phi$ \cite{Tomberg:2024evi}. It is possibly time dependent. In general, the noise amplitude should be ``pre-computed" through the solutions of the relevant $Q_k$ modes.    

We will use these equations to answer the following questions: 
\begin{itemize}
	\item[a)]
	{\it What dynamical equation for $\Pi$ is implied by (\ref{HJ-Stoch}) and (\ref{noise1}) and how does it compare to the one in the commonly used phase space approach? }
	\item[b)] {\it What is the dynamical evolution of perturbations on long wavelengths if HJ is an accurate description? }
\end{itemize}
To obtain answers, we write the displacement $\Delta\phi$ for a short timestep $\Delta t$   
\beq\label{Delta-Stoch}
\Delta\phi = -2\frac{dH}{d\phi} N \Delta t + \xi_\phi N\Delta t
\,,\quad
\eeq  
and compute the corresponding change in $\Pi$. We note that when dealing with stochastic differential equations we can effectively set $\left(\xi dt\right)^2 
\rightarrow \mathcal{A} dt$, which is a heuristic way to obtain  It$\hat{o}$'s formula given \eqref{noise1}.  Therefore, to obtain the $\Pi$ differential equation given the drift+stochastic displacement $\Delta\phi$ we need to expand $\Pi$ to order $(\Delta\phi)^2$ and keep only terms linear in $\Delta t$.   

We use e-folds as the time variable ($\alpha=\ln a$)
\begin{equation}
	N\Delta t = \frac{1}{H}\Delta\alpha\,.
\end{equation}
Writing $\Pi \equiv H\pi
=Hd\phi/d\alpha$, we have 
\beq
\Delta\phi = -2\frac{d\ln H}{d\phi}  \Delta \alpha + \frac{\xi_\phi}{H} \Delta \alpha  = \pi \Delta\alpha + \frac{\xi_\phi}{H} \Delta \alpha 
	\,,\quad \pi = -\sqrt{2\epsilon}
	\,,\quad \epsilon=\frac12\pi^2 = 3 - \frac{V}{H^2} 
\,,\quad
\eeq 
and obtain that for a small field displacement $\Delta\phi$ the corresponding $\pi$ displacement is:
\beq
\Delta \pi = -2\frac{{\rm d}^2\ln H}{{\rm d}\phi^2} \Delta\phi
- \frac{{\rm d}^3\ln H}{{\rm d}\phi^3}\Delta\phi^2 
\,.\quad
\eeq
As mentioned above, for the stochastic dynamics we should set the stochastic part of $\Delta\phi^2 \rightarrow \frac{\mathcal{A}}{H}\Delta\alpha
$ 
for the stochastic part of the field displacement and we have
%
%
\beq
\label{Pi-stoch3}
	\Delta \pi = \frac{\epsilon_2}{2}\pi \Delta\alpha 
	+ \frac{\epsilon_2\epsilon_3}{4}\frac{\mathcal{A}}{H} \Delta \alpha + \frac{\epsilon_2}{2}\frac{\xi_\phi}{H}\Delta\alpha
\,,\quad
\eeq
where we used
\beq\label{eq:eps2}
\frac{d^2\ln H}{d\phi^2} = \frac{1}{2}\frac{V}{H^2}+\frac{1}{2}\frac{V'}{H^2\pi} 
= -\frac{\epsilon_2}{4}
\,,\quad
\eeq
and that
\beq
\epsilon_2' = \epsilon_2\epsilon_3
\,.
\eeq
The first term in (\ref{Pi-stoch3}) is the standard drift term of the momentum evolution equation when the number of e-folds is taken as the time variable. The second term is an additional drift term that arises due to the fact that $H(\phi)$ is a non-linear function of the stochastic variable $\phi$ - see 
{\it e.g.}~\cite{Gardiner1994}. It will not be relevant as long as $ \sqrt{8\epsilon_1}/\epsilon_3\gg \mathcal{A}/H\simeq \left(H/2\pi\right)^2$. This inequality should be valid for inflation taking place at sufficiently sub-Planckian energies and for slow-roll models, where the l.h.s. is a large number. For our USR model it has a minimum at $\left[\sqrt{8\epsilon_1}/\epsilon_3\right]_{\rm min} \simeq 0.092$. We therefore neglect this stochastically induced drift term for subplanckian inflation. 

We thus arrive at the momentum evolution equation as implied by the (stochastic) HJ \eqref{eq:evol-HJ}
\beq\label{eq:HJ-induced-momentum}
\frac{d\pi}{d\alpha} +\frac{V}{H^2}\pi+\frac{V'}{H^2}  =\frac{\epsilon_2}{2}\frac{\xi_\phi}{H}
\,.\quad
\eeq
The l.h.s.~of \eqref{eq:HJ-induced-momentum} is the expected dynamical equation for the momentum which further shows that, for long wavelength fields, \begin{equation}\label{eq:adiabatic}
\Delta\pi = \frac{\epsilon_2}{2}\Delta\phi\,.
\end{equation}
Adiabatic perturbations, which are equivalent to time displacements, satisfy relation \eqref{eq:adiabatic}. The final term in \eqref{eq:HJ-induced-momentum} is the stochastic noise for the momentum which shows that if (\ref{HJ-Stoch}) is used as the dynamical equation, the implied stochastic noise for the momentum is related to that of the field via 
\beq\label{LW velocity}
\xi_\pi = \frac{\epsilon_2}{2}\xi_\phi\,. 
\eeq
Thus, the momentum and field stochastic forces are fully correlated and with a specific relation between their amplitudes.  Of course, this is nothing but the noise one gets in the standard phase-space formulation if perturbations are in their adiabatic mode: since
\begin{equation}
	\xi_\pi = \xi_\phi\frac{\partial_\alpha\delta{\phi}_k}{\delta\phi_k}\Big|_{k=k_\sigma}
\end{equation}   
Eq.~\eqref{LW velocity} will hold if
\beq\label{eq:LW_modes2}
\partial_\alpha\delta{\phi}_k = \frac{\epsilon_2}{2} \delta\phi_k
\,,\quad
\eeq
when filtered into the long wavelength sector (bounded from above in $k$-space by $k<k_\sigma$), see e.g.~\cite{Tomberg:2022mkt}. We have thus shown that, as long as the modes satisfy \eqref{eq:long_modes}, the stochastic equation for the momentum is a superfluous equation in the HJ formalism as it is implied by the first order HJ stochastic equation \eqref{HJ-Stoch}. We now examine the validity of \eqref{eq:LW_modes2} in our USR model.


%

\section{Perturbation Modes, HJ Attractors and the Conveyor Belt}
\label{sec:HJ+conveyor}

\begin{figure}[t]
	\centering
	\includegraphics[width=0.6\textwidth]{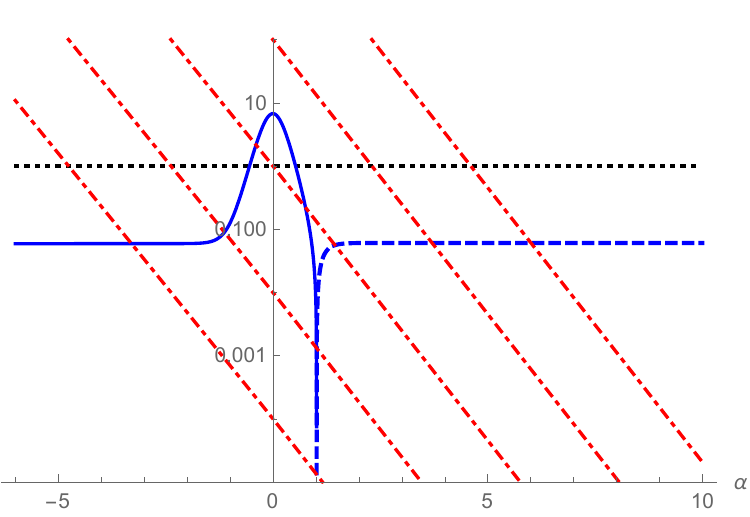}
	\caption{The "mass term" $m^2(\alpha)$ in~\eqref{mass term of Q} for $\epsilon_0 = 0.03$ and $\eta_0=-0.02$ is denoted by the blue line (the dashed part denotes $m^2<0$) as a function of e-folds, which run left to right. The red dash-dotted lines represent $(k/aH)^2$ for 5 modes $(k,\,10 k,\,10^2 k,\,10^3 k ,\,10^4 k)$. The modes become super-Hubble $(k=aH)$ when the black dashed line is crossed.}
	\label{fig:MassTerm}
\end{figure}

We solve the mode equation \eqref{Field eq 1} for a set of fixed modes $\delta\Phi(\alpha,\mathbf{k})$ to illustrate how different wavenumbers behave during the USR phase and check the validity of HJ by comparing the mode behaviour with the HJ prediction \eqref{eq:adiabatic} and \eqref{eq:LW_modes2}. Fig.~\ref{fig:MassTerm} shows the mass term of our model \eqref{mass term of Q} (blue line) and 5 modes $(k,10k,\ldots ,\,10^4 k)$ (red dash-dotted lines) spanning 4 decades in $k$-space that cross the Hubble radius ($k=aH$ - black dashed line) at distinctive times during the time evolution of our model. The first two modes cross during the pre-USR phase and are super-Hubble when the mass term  becomes relevant for them. The third mode exits during the transition and feels the effect of the mass term while still sub-Hubble. The fourth and fifth modes exit after the transition. We call them modes $1-5$ in our discussion below. 

\begin{figure}[t]
	\centering
	\includegraphics[width=0.49\textwidth]{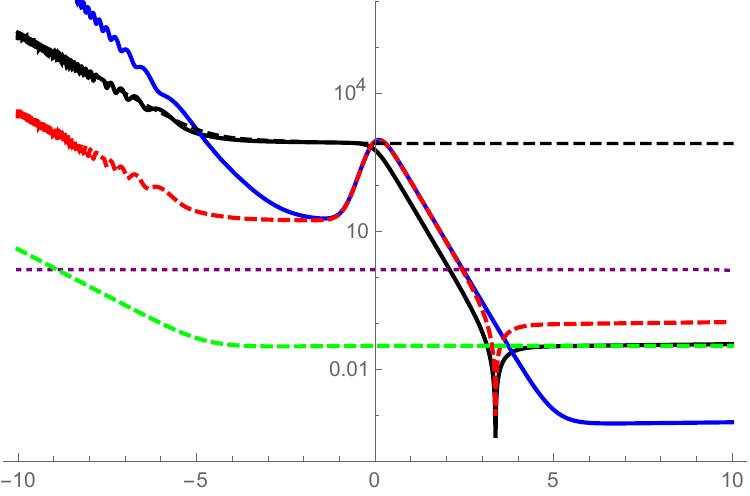}
	\includegraphics[width=0.49\textwidth]{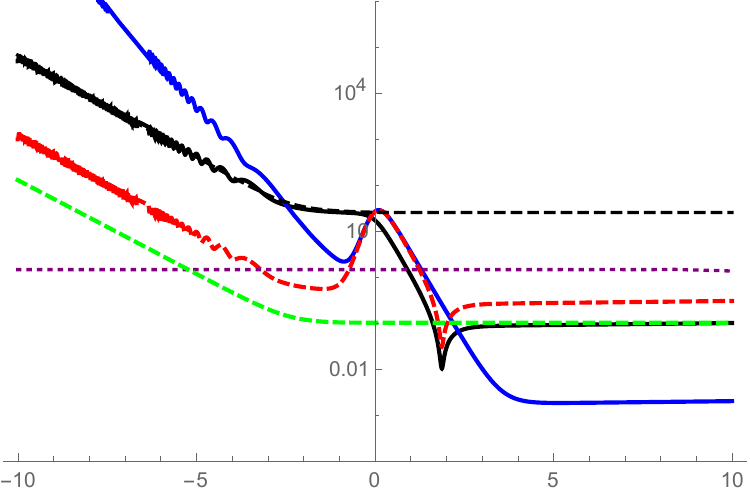}
	\caption{Modes 1 and 2. The amplitude of the solution to (\ref{Field eq 1}) is shown as the black solid line, with the dashed black line denoting the expected de Sitter result with $H=H_0$ if no USR were to follow the modes' Hubble exit. During USR, the field perturbation decays, settling to a constant lower amplitude. The green dashed  line is our prediction for the final mode amplitude (\ref{USRL field decay}), applicable to modes that have exited before and during the transition to USR. The blue line is the field velocity and the red dashed line is the HJ prediction \eqref{eq:long_modes}, tracing the decaying part of the velocity well until the subdominant constant amplitude solution is reached. The constant dashed line is computed from the numerical evaluation of the modes' Wronskian, indicating that the numerical solution is accurately obtained.}
	\label{fig:HJQModes1}
\end{figure}

\begin{figure}[t]
	\centering
	\includegraphics[width=0.6\textwidth]{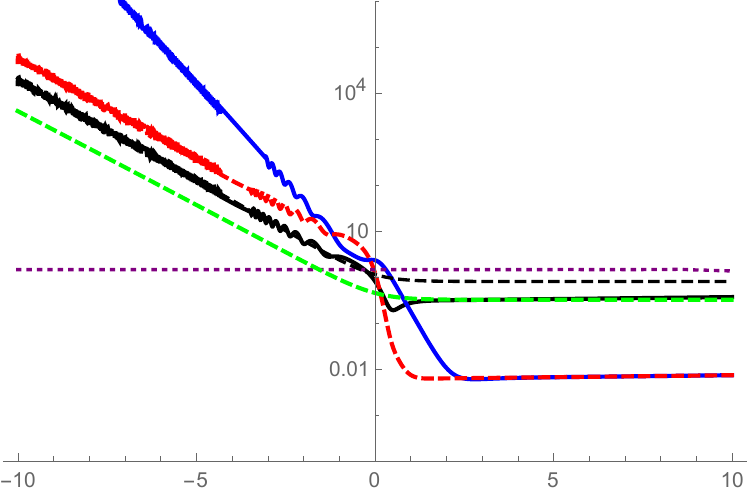}
	\caption{Mode 3. The amplitude of the solution to 
    (\ref{Field eq 1}) is shown as the black solid line, with the dashed black line denoting the expected de Sitter result with $H=H_0$. The green dashed line is our prediction for the final mode amplitude (\ref{USRL field decay}), applicable to modes that have exited before and during the transition to USR. The blue line is the field velocity. The red dashed line is now the HJ prediction \eqref{eq:long_modes} but in the new HJ branch with $\tilde{\epsilon}_2$ instead of $\epsilon_2$, correctly capturing the final mode velocity. The constant dashed line is computed from the numerical evaluation of the modes' Wronskian, indicating that the numerical solution is accurately obtained. }
	\label{fig:HJQModes2}
\end{figure}

\begin{figure}[t]
	\centering

	\includegraphics[width=0.49\textwidth]{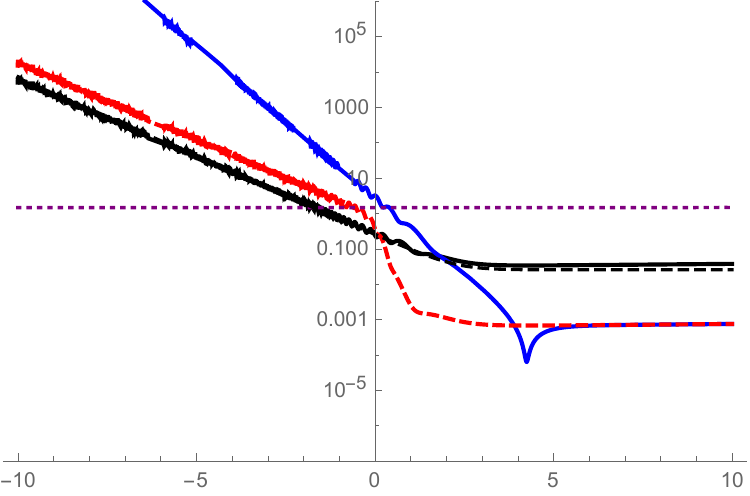}
	\includegraphics[width=0.49\textwidth]{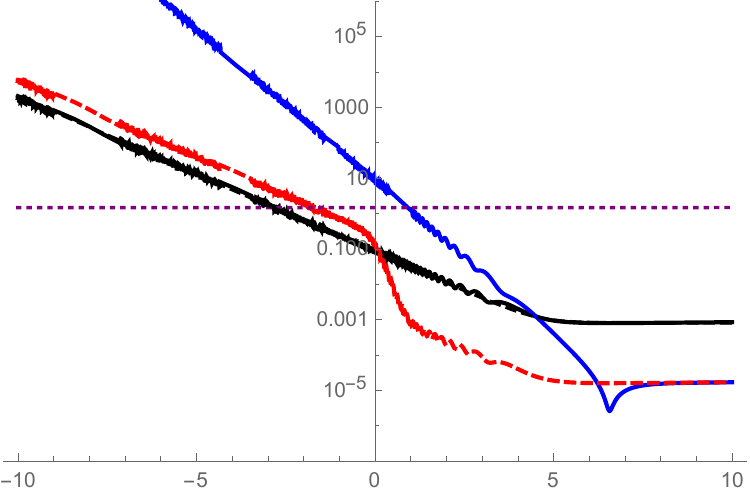}
	\caption{Modes 4 and 5. The amplitude of the solution to \ref{Field eq 1} is shown as the black solid line, with the dashed black line denoting the expected de Sitter result with $H=H_0$. The blue line is the field velocity. The red dashed line is the HJ prediction \eqref{eq:long_modes} but again in the new HJ branch with $\tilde{\epsilon}_2$ instead of $\epsilon_2$, capturing the post-USR part of the mode velocity. The constant dashed line is computed from the numerical evaluation of the modes' Wronskian, indicating that the numerical solution is accurately obtained. }
	\label{fig:HJQModes3}
\end{figure}

We plot the behaviour of these illustrative 5 modes in Fig.~\ref{fig:HJQModes1} (modes 1 and 2 which exit prior to the USR transition), Fig.~\ref{fig:HJQModes2} (mode 3 which exits during the USR transition) and Fig.~\ref{fig:HJQModes3} (modes 4 and 5 which exit after the USR transition). The black and blue lines are the mode amplitude and mode velocity respectively, and the red dashed line is the mode amplitude-velocity relation \eqref{eq:long_modes}, predicted by HJ theory. The horizontal dotted line shows the numerically computed Wronskian of the modes, which provides a diagnostic of the accuracy of the numerical solution. 

Modes $1$ and $2$ (exit times $\alpha \simeq -5$ and $\alpha \simeq -2$, respectively) settle into the expected approximate de Sitter amplitude (black dashed line) and at the USR transition start decaying as $\delta\Phi\simeq a^{-3}$. This is precisely what HJ predicts \cite{Prokopec:2019srf}\footnote{Such decaying behaviour is also evident in the full Numerical Relativity solution for a USR potential of \cite{Launay:2025kef}.} and this can be seen more quantitatively by the overlap of the numerically computed velocity with relation \eqref{eq:long_modes}- overlap of blue with red-dashed lines. A difference from the long wavelength HJ expectation is that the decay does not last indefinitely and the mode settles to a subdominant constant amplitude.

Mode 3 (exit at $\alpha=0$) is displayed in Fig.~\ref{fig:HJQModes2} and exits during the USR phase. The mode settles to a slightly reduced amplitude compared to the de Sitter expectation. The relation \eqref{eq:long_modes} with the USR $\epsilon_2 = -6+2\eta_0$ fails. This appears consistent with the comments in \cite{Tomberg:2022mkt} and the computation in \cite{Artigas:2025nbm} which show that terms neglected in HJ can be relevant during USR. However, using a {\it new} HJ branch { by shifting to the slow-roll branch of the potential \eqref{potential around minimum} which also gives the dual of $\epsilon_2 = -6+2\eta_0$},
\begin{equation}
    \epsilon_2 \rightarrow \tilde{\epsilon}_2 = -2\eta_0
\,,\quad
\end{equation}
gives the correct final velocity for the mode -- this new HJ prediction is displayed by the red dashed line in Fig.~\ref{fig:HJQModes2} which overlaps with the numerically determined mode velocity (blue line) for $\alpha \gtrsim 2$.\footnote{It is worth noting here that shifting to $\tilde{\epsilon}_2$ also matches the final velocity of modes 1 and 2.} The precise details behind such a jump between HJ branches still need to be investigated. The observation that the non-isotropic part of the momentum constraint (which is dropped in the HJ formalism) can become important in USR \cite{Tomberg:2022mkt, Artigas:2025nbm}, is likely of relevance for explaining in detail the jump to another HJ branch. We emphasize however, that {\it this does not invalidate the conveyor-belt framework}, which is supported by our current numerical results. Instead, it may offer a possible clarification for the precise dynamics via which the transition between HJ branches occurs.

The final amplitude for modes 1 -- 3 can be fit well by an expression of the form:
\begin{equation}
		\frac{\delta\Phi(\alpha\rightarrow \infty)}{\delta\Phi_{\rm dS}}
		\simeq \mathcal{B} \left(\frac{k}{H}\right)^{\!2}
		\,.
		\label{USRL field decay}
	\end{equation}
This is shown by the green dashed line in Figs.~\ref{fig:HJQModes1} and \ref{fig:HJQModes2}. We present an analytical argument for the $(k/H)^2$ dependence of the residual amplitude in Appendix \ref{app:k_sq-terms}, with the precise value of the coefficient $\mathcal{B}$ determined numerically and depending on the model parameters $\epsilon_0$ and $\eta_0$; the value $\mathcal{B}=2/5$ has been obtained from our numerical solutions and used in the figures.  

Modes 4 and 5 are shown in Fig.~\ref{fig:HJQModes3} and exit after the USR transition (exit times $\alpha \simeq 2$ and $\alpha \simeq 5$ respectively). The mode amplitude settles in a new regime with a small residual temporal variation due to the final non-zero value of:
\begin{equation}
	m^2(\alpha) \rightarrow \eta_0\left(3 -\eta_0\right)
	\,.
	\label{eq:final mass}
\end{equation}
The final value of the velocity again matches the new $\tilde{\epsilon}_2$ HJ branch as can be seen by the agreement of the blue and dashed red lines. The final mode amplitude agrees with an approximately de Sitter, { slow-roll spacetime}, corresponding to the shifted HJ branch, and formula \eqref{USRL field decay} is no longer applicable.

%
%
%
%

\section{Discussion}
\label{Discussion}

The above numerical results allow us to reach some general conclusions about the use of HJ to describe USR:

\begin{itemize}
    \item The HJ stochastic equation \eqref{HJ-Stoch} (where $\mathcal{A}$ should be consistently determined from the mode functions) gives a picture of the dynamics equivalent to the phase space one as long as the coarse grained noise is computed on the overlap of the blue and red curves in fig.~\ref{fig:HJQModes1}. More generally, modes that have become super-horizon prior to the USR transition, experience a decay while carried through the transition, as predicted by HJ - see \cite{Prokopec:2019srf} for the simpler case $V=V_0$.    
	
    \item The HJ analysis of \citep{Prokopec:2019srf} predicted a complete decay of all inflaton perturbations introduced prior to the transition: in the $V=V_0$ case $\phi(\alpha,\mathbf{x}) \rightarrow \phi_0$ $\forall \, \mathbf{x}$. In reality there is a residual freeze-out at subdominant levels given by \eqref{USRL field decay}. The freeze-out amplitude should only have a small effect since diffusion from modes that exit later would dominate over this residual fluctuation. In the context of the constant $V=V_0$ potential, this can can be thought as promoting $\phi_0$ of \cite{Prokopec:2019srf} to have an $\mathbf{x}$ dependence; this was implemented in \cite{Wilkins:2023asp}. 

    \item Modes exiting during and after the USR transition are no longer described by the HJ branch with $\epsilon_2\simeq -6$. However,  
    \beq
    \partial_\alpha\delta{\Phi}_k = \frac{\tilde{\epsilon}_2}{2} \delta\Phi_k
\,,\quad
    \eeq   
    does very well in describing the velocity after the USR transition. The new HJ branch corresponds to $\tilde{\epsilon}_1(\alpha) = -\epsilon_1(-\alpha)$ which can be interpreted for $\eta_0<0$ as an HJ solution where the field is slow-rolling down the potential slope, moving away from $\phi_0$. This is precisely the conveyor belt idea, put forward in \cite{Prokopec:2019srf}, whereby fluctuations stemming from sub-Hubble scales necessitate the transition between otherwise disjoint solutions of the HJ equation. In our example, the analogous situation in the conveyor belt of a slide on $V=V_0$ was the transition from the $\Pi\neq 0$ solution to the $\Pi=0$ solution.   

    \item{ A long wavelength solution with $\epsilon_2 \rightarrow -6$ does not appear to represent a physical asymptotic future spacetime. From \eqref{eq:eps2} we see that such a value results from the field transversing a part of the potential $V$ {\it faster} than the attractor velocity set by the slope $V'$,  (but with its kinetic energy still being subdominant to its potential energy).\footnote{ Hence the term USR is arguably a misnomer.} The decay of long wavelength inhomogeneities during this USR phase makes spatial gradient terms important even in the IR, when the amplitude of the long wavelength fluctuations has decayed sufficiently, at which point Eq.~(\ref{eq:LW_modes2}) breaks down. It appears that such gradient terms then force the system to eventually shift to the attractor where $\pi\simeq -V'/(\sqrt{3}V)$. Once this occurs, the new attractor emerges as a description of the long-wavelength evolution and the gradient terms may again be neglected.\footnote{ A. Starobinsky was already aware that HJ attractors are the general attractors for inflationary backgrounds -- personal communication with T. Prokopec.} Although this picture explains our observations regarding the conveyor belt, a more detailed analysis would be required for a rigorous demonstration of such a conjecture. In this respect, a detailed comparison with the results of \cite{Artigas:2025nbm} which demonstrate the relevance during USR of terms neglected in the HJ formulation \eqref{HJ} and \eqref{momentum-def} may prove illuminating.}
	
\end{itemize}

It is important to emphasize that in using the HJ attractor to formulate a stochastic HJ equation, only one dynamical degree of freedom is involved, supplied by the energy and momentum conservation laws. The formalism then provides an accurate description of super-Hubble dynamics as long as spatial gradients are negligible. The USR model discussed in this work represents one instance in which the role of spatial gradients can be important for determining the final amplitude of the field perturbation. It is therefore required to formulate an approach for when spatial gradients become relevant in the stochastic HJ formalism { and the results of \cite{Artigas:2025nbm} appear relevant for this}. The figures and the computation in appendix \ref{app:k_sq-terms}, involving the mixing of the two linearly independent perturbation modes, offer some clues but we leave discussion of general conditions for future work. However, regardless of any precise such conditions, it appears from our numerical investigation that the main effect of gradient terms is the eventual transition to different HJ branches { of the same underlying potential. This is therefore an example where knowledge of the geometric slow roll parameters $\epsilon_i$ does not completely specify the evolution of the system. 

These conclusions align with the concept of the conveyor belt} put forward in \cite{Prokopec:2019srf} which we have verified here via the numerical solution of the gauge invariant field perturbation for our USR model. The involvement of both perturbation modes during USR was already emphasized in \cite{Artigas:2025nbm} via the importance of the non-adiabatic pressure perturbation. However, their criticism of the use of HJ for USR, with HJ described as relying on a { single, global HJ attractor}, is misplaced: The conveyor belt of \cite{Prokopec:2019srf} explicitly makes use of both long-wavelength dynamical behaviours allowed by the potential; these correspond to the two linearly independent perturbation modes in linear theory and mix in USR. While we agree that non-adiabatic perturbations in the momentum constraint play an important role in USR, the effect they have on perturbations appears to be already encoded { qualitatively} in the HJ conveyor belt stochastic formalism of [1].  

We end by recalling that the stochastic,  Langevin type equation discussed in Sec.~\ref{sec:Stoch-HJ} refers to the scalar field $\phi$. Such equations are commonly used to describe long wavelength scalar dynamics in inflation, and are expected to describe non-linear, inhomogeneous field configurations, beyond linear perturbation theory. However, a further first passage time analysis (the stochastic $\Delta N$ formalism) \cite{Vennin:2015hra} would be required to obtain the curvature perturbation which ends up being the observable produced by inflation.

\section*{Acknowledgments}
TP was supported by the D-ITP consortium, a program of the Netherlands Organization for Scientific Research (NWO) that is funded by the Dutch Ministry of Education, Culture and Science (OCW).

\appendix

\section{Commonly used geometric slow-roll parameters}\label{Appendix A}

Here we discuss two commonly used sets of geometric slow-roll 
parameters. Both share the principal slow-roll parameter $\epsilon$,
which in (HJ) attractor models of inflation (which is what 
we assume here) can be expressed in terms of the Hamiltonian constraint,
\begin{equation}
\epsilon(\phi) = \frac{\pi^2}{2} 
      = 2\left(\frac{{\rm d}\ln(H[\phi)]}{{\rm d}\alpha}\right)^2
        = 3 - \frac{V(\phi)}{H^2(\phi)}
\,,
\label{appx: principal eps}
\end{equation}
where $\alpha = \ln(a)$ is a time variable parametrizing e-folding
 of inflation (in an attractor regime of inflation, there is no reason to introduce a special time parameter, as it can be characterized by the inflaton $\phi$) From~(\ref{appx: principal eps}) we see that 
 $\epsilon$ maiximizes at 3 in kination, in which $V=0$, and 
the de Sitter space is obtained when $V=3H^2$, which are familiar limits. 

The first set is defined in terms of the derivatives of the principal slow 
roll parameter, 
\begin{equation}
\epsilon_{i+1} = \frac{{\rm d}\ln[\epsilon_i]}{{\rm d}\alpha}
\,,\quad (i=1,2,\cdots)\quad {\rm and} \;\; \epsilon_1\equiv \epsilon
\,.
\label{first set of slow-roll}
\end{equation}
The second set, used mostly in this work, is defined by
\begin{equation}
\epsilon = \epsilon_1
\,,\qquad 
\eta = -3-\frac{1}{\pi}\frac{V'(\phi)}{H^2(\phi)}
\,,
\label{appx: second set of slow-roll}
\end{equation}
where $\pi = {\rm d}\phi/{\rm d}\alpha = -\sqrt{2\epsilon}$.

Upon taking a derivative of~(\ref{appx: principal eps})
with respect to $\alpha$, and taking 
into account that $\epsilon_1=\epsilon$, one obtains, 
\begin{equation}
\epsilon_2 = 2(\epsilon+\eta) \;\; \;\leftrightarrow \; \;\;
     \eta = -\epsilon_1 + \frac{\epsilon_2}{2}
\,.
\label{relating slow-roll parameters 2}
\end{equation}
It is also worth mentioning the next order in slow-roll, 
\begin{equation}
\epsilon_2 = \frac{\epsilon'}{\epsilon}
\,,\quad \epsilon_3 = \frac{(\epsilon+\eta)'}{\epsilon+\eta}
\;\;\;\leftrightarrow \; \;\;
\epsilon' = \epsilon_1\epsilon_2
\,,\quad
(\epsilon+\eta)' = \frac12 \epsilon_2\epsilon_3
\,.
\label{relating slow-roll parameters 3}
\end{equation}

Classical perturbations obey second order differential equations,
which at the linear order in perturbation is known 
as the Mukhanov equation,
\begin{equation}
\Big(\partial_\tau^2 + k^2 - \frac{z''}{z}\Big)Q(\tau,\vec k) = 0
\,,
\label{Mukhanov eq2}
\end{equation}
where $\tau$ is conformal time ($a{\rm d}\tau=N{\rm d}t$), $z'=\partial_\tau z$, 
$Q = a\delta \phi-z\psi$ 
is the gauge invariant Mukhanov-Sasaki field and 
$z = \sqrt{2\epsilon}a$ (in terms of which the curvature perturbation is given by 
${\cal R}=-Q/z$).
To convert this equation to the formalism used here, first notice 
that the times are related as, $a(\tau) {\rm d}\tau = N{\rm d}t$,
from which one infers, 
$\partial_\tau = \frac{a}{N}\partial_t = H{\rm e}^\alpha\partial_\alpha$.
This then implies for $z=\sqrt{2\epsilon}a$, 
\begin{equation}
z'(\tau) = aH\left(1+\frac{\epsilon}{2}\right)z
\label{appx: z prime}
\end{equation}
and 
\begin{eqnarray}
\frac{z''}{z} &=& (aH)^2\left[2-\epsilon_1+\frac32\epsilon_2
-\frac12\epsilon_1\epsilon_2
+\frac14\epsilon_2^2
+\frac12\epsilon_2\epsilon_3\right]\\
&=& (aH)^2\left[2+2\epsilon+3\eta
+\frac12\epsilon' + \eta'
+(\epsilon+\eta)^2\right]
\,.
\label{appx: z'' over z}
\end{eqnarray}
Inseting these results and 
\begin{equation}
\partial_\tau^2 
 = (Ha)^2\big[(1-\epsilon)\partial_\alpha +\partial_\alpha^2\big]
 \,,
\label{appx: conversion d eta}
\end{equation}
into 
Eq.~(\ref{Mukhanov eq}) one obtains, 
\begin{eqnarray}
\left\{\partial_\alpha^2 + (1-\epsilon_1)\partial_\alpha+ \frac{k^2}{(aH)^2}
 - \left[2-\epsilon_1+\frac32\epsilon_2
-\frac12\epsilon_1\epsilon_2
+\frac14\epsilon_2^2
+\frac12\epsilon_2\epsilon_3\right]
\right\}Q(\alpha;\vec k) &=& 0
\qquad
\label{Mukhanov eq 2}\\
\left\{\partial_\alpha^2 + (1-\epsilon)\partial_\alpha+ \frac{k^2}{(aH)^2}
 -\left[2+2\epsilon+3\eta
+\frac12\epsilon' + \eta'
+(\epsilon+\eta)^2\right]
\right\}Q(\alpha;\vec k) &=& 0
\,.\qquad\;
\label{Mukhanov eq 3}
\end{eqnarray}
For the (gauge invariant) field perturbation, ${\delta\Phi}=Q/a$, one gets,  
\begin{eqnarray}
\left\{\partial_\alpha^2 + (3-\epsilon_1)\partial_\alpha+ \frac{k^2}{(aH)^2}
 -\frac32\epsilon_2
+\frac12\epsilon_1\epsilon_2
-\frac14\epsilon_2^2
-\frac12\epsilon_2\epsilon_3
\right\}{\delta\Phi}(\alpha;\vec k) &=& 0
\qquad
\label{Field eq 12}\\
\left\{\partial_\alpha^2 + (3-\epsilon)\partial_\alpha+ \frac{k^2}{(aH)^2}
 -3(\epsilon+\eta)
-\frac12\epsilon' - \eta'
-(\epsilon+\eta)^2
\right\}{\delta\Phi}(\alpha;\vec k) &=& 0
\,.
\label{Field eq 2}
\end{eqnarray}
%


\section{Analytical estimate for the residual amplitude in USR}\label{app:k_sq-terms}

From figure~\ref{fig:MassTerm} we see that the mass term $m^2$ peaks 
at $m^2(0)\simeq 3(9\!-\!\epsilon_0)$ at $\alpha=0$, and it decays thereafter as $m^2\simeq 3(9\!-\!\epsilon_0){\rm e}^{-6\alpha}$, until it reaches a constant value $m^2\propto\eta_0$.
During that time both the field $\delta\Phi$ and its momentum $\pi$ 
decay as $\delta\Phi\propto {\rm e}^{-3\alpha}$, 
the subsequent evolution, during which 
$\frac{k^2}{H^2}{\rm e}^{-2\alpha}\gg m^2(\alpha)$,
 can be well approximated by that of de Sitter sub-Hubble massless modes, 
 for which the normalized positive and negative frequency 
 mode functions are (recall 
 $\frac{1}{a} = {\rm e}^{-\alpha} = -H\eta$),
 \begin{eqnarray}
 \phi(\alpha,k)  &=& \frac{H}{\sqrt{2k^3}}
 \left(1\!-\!i\frac{k}{H}{\rm e}^{-\alpha}\right)
 \exp\left(i\frac{k}{H}{\rm e}^{-\alpha}\right)
\,,\nonumber\\
 \phi^*(\alpha,k)  &=& \frac{H}{\sqrt{2k^3}}
 \left(1\!+\!i\frac{k}{H}{\rm e}^{-\alpha}\right)
 \exp\left(-i\frac{k}{H}{\rm e}^{-\alpha}\right)
 \,.
 \label{appx: exact dS mode functions}
\end{eqnarray}
From these modes one can construct a general solution, 
\begin{equation}
\delta\Phi = \beta(k) \phi(\alpha,k) + \gamma(k) \phi^*(\alpha,k)
\,,
\label{general field}
\end{equation}
where $\beta(k)$ an $\gamma(k)$ are Bogolyubov coefficients satisfying,
$|\beta|^2-|\gamma|^2=1$, which can be determined by requiring  
that the relation $\partial_\alpha\delta\Phi(0,k)\simeq -3\delta\Phi(0,k)$
is satisfied at $\alpha=0$. A simple calculation shows that
the Bogolyubov coefficients are, 
\begin{equation}
\beta(k)=\gamma^*(k) =
\frac{3\phi^*+\partial_\alpha\phi^*}{W[\phi,\phi^*]}\delta\Phi_{dS}
= \frac{H^2}{\sqrt{2k^3}}\bigg[3\frac{k}{H}-i\Big(3
-\frac{k^2}{H^2}\Big)\bigg]e^{-i\frac{k}{H}}\delta\Phi_{\rm dS}
\,.
\label{appx: beta and gamma}
\end{equation}
These Bogolyubov coefficients are approximate, in that 
they satisfy $|\beta|^2-|\gamma|^2=0$, which is due to the fact 
that the equation $\partial_\alpha\delta\Phi(0,k)\simeq -3\delta\Phi(0,k)$ is not exact. The suppression of the modes in USR is obtained by calculating the ratio,
\begin{equation}
\frac{\delta \Phi(\infty)}{\delta\Phi(0)}
\simeq \frac{\delta \Phi(\infty)}{\delta\Phi_{dS}}
     \simeq -\frac{1}{15}\left(\frac{k}{H}\right)^2
\,,
\label{appx: k2 suppression}
\end{equation}
and it can be used to compare with numerical results.
Numerical integration confirms the scaling 
$|\delta \Phi(\infty)/\delta\Phi_{dS}|\sim k^2/H^2$,
but the actual coefficient is somewhat larger than $1/15$ and depends on the model parameters, here $\epsilon_0$ and $\eta_0$.

%
%

      
\end{document}